\documentclass[12pt,onecolumn,draftcls]{IEEEtran}
\usepackage{graphicx,amssymb,amsmath}
\usepackage{multicol}
\usepackage[noadjust]{cite}
\usepackage{setspace}
\usepackage{stfloats}
\usepackage{midfloat}
\usepackage[normal]{threeparttable}
\usepackage{amsthm}
\usepackage{cuted}
\usepackage{cases,subeqnarray}
\usepackage{bm,multirow,bigstrut}
\usepackage{textcomp}
\usepackage{latexsym,bm}
\usepackage{booktabs,changebar}
\usepackage{xcolor}
\usepackage{extarrows}
\usepackage{tablefootnote}
\theoremstyle{plain}
\newtheorem{thm}{Theorem}
\newtheorem{lemm}{Lemma}
\theoremstyle{plain}

\IEEEoverridecommandlockouts
\begin{document}
\title{On High-Order Capacity Statistics of Spectrum Aggregation Systems over $\kappa$-$\mu$ and $\kappa$-$\mu$ shadowed Fading Channels}

\author{Jiayi~Zhang,~\IEEEmembership{Member,~IEEE,}
        Xiaoyu~Chen,
        Kostas~P.~Peppas,~\IEEEmembership{Senior Member,~IEEE,}
        Xu~Li
        and~Ying~Liu
\thanks{
This work was supported in part by the National Natural Science Foundation of China (Grant Nos. 61601020 and 61371068), the Fundamental Research Funds for the Central Universities (Grant Nos. 2016RC013, 2016JBZ003 and 2014JBZ002). 
}%
\thanks{J. Zhang, X. Chen, X. Li, and Y. Liu are with the School of Electronics and Information Engineering, Beijing Jiaotong University, Beijing 100044, P. R. China (e-mail: jiayizhang@bjtu.edu.cn).}
\thanks{K. P. Peppas is the Department of Informatics and Telecommunications, University of Peloponnese, 22100 Tripoli, Greece. (e-mail: peppas@uop.gr).}
}
\maketitle

\begin{abstract}
The frequency scarcity imposed by fast growing demand for mobile data service requires promising spectrum aggregation systems. The so-called higher-order statistics (HOS) of the channel capacity is a suitable metric on the system performance. While prior relevant works have improved our knowledge on the HOS characterization of spectrum aggregation systems, an analytical framework encompassing generalized fading models of interest is not yet available. In this paper, we pursue a detailed HOS analysis of $\kappa$-$\mu$ and $\kappa$-$\mu$ shadowed fading channels by deriving novel and exact expressions. Furthermore, the simplified HOS expressions for the asymptotically low and high signal-to-noise regimes are derived. Several important statistical measures, such as amount of fading, amount of dispersion, reliability, skewness, and kurtosis, are obtained by using the HOS results. More importantly, the useful implications of system and fading parameters on spectrum aggregation systems are investigated for channel selection. Finally, all derived expressions are validated via Monte-Carlo simulations.
\end{abstract}

\begin{IEEEkeywords}
Spectrum aggregation, Higher-order statistics, $\kappa$-$\mu$, $\kappa$-$\mu$ shadowed.
\end{IEEEkeywords}

\IEEEpeerreviewmaketitle
\section{Introduction}
With the constantly growing mobile data demand for future wireless communication systems, i.e., 5th Generation (5G), it becomes more and more difficult to allocate a wide and contiguous frequency band to each user equipment (UE) and base station (BS). This has brought about increasing scarcity in available radio spectrum. To address these issues, the promising spectrum aggregation technique has been received much attention recently \cite{bogucka2015dynamic,zhang2015lte}. Spectrum aggregation refers to obtaining larger amounts of radio resource by aggregating possible spectrum resources that lie in non-adjacent frequency bands. As a successful application of the spectrum aggregation, the carrier aggregation (CA) technology has been proposed in Long-Term-Evolution Advanced (LTE-A) standard, increasing the usable spectrum by aggregating resource blocks (RBs) either within a given band or in different frequency bands \cite{zhang2015lte}. In order to achieve a successful spectrum aggregation, the maximum dispersion in the channel capacity should be calibrated to leverage a reliable transmission \cite{yilmaz2014computation}.

In this context, the typical metric for performance evaluation has been the higher-order statistics (HOS) of the channel capacity, which can fully explore the reliability of the signal transmission in spectrum aggregation systems. As an useful tool, the HOS can effectively describe the channel capacity dispersion induced by the heterogeneity that inherently exists in spectrum aggregation systems \cite{simon2005digital}. Moreover, fruitful insights into the implications of the spectrum aggregation on the transmission reliability can be extracted by deriving HOS of the channel capacity. Despite its importance, however, the HOS of the channel capacity received relatively little attention in the literature, due in part to the intractability of its analysis. A number of prior works have investigated the HOS of the channel capacity of different wireless systems over several flat fading channels \cite{yilmaz2012novel,yilmaz2014computation,yilmaz2012computation,sagias2011higher}. For example, a generic framework for the asymptotic HOS of the channel capacity over independent and identically distributed (i.i.d.) Nakagami-$m$ fading channels was provided in \cite{yilmaz2012novel}. The authors in \cite{yilmaz2014computation} investigated the HOS of the channel capacity for amplify-and-forward (AF) multihop systems over gamma and generalized gamma fading channels. %
{In addition, an MGF-based approach for the HOS of the channel capacity for $L$-branch MRC receivers has been proposed in \cite{yilmaz2012computation} with an example application of correlated Nakagami-$m$ fading channels.} Finally, \cite{sagias2011higher} presented the HOS of the channel capacity for several diversity receivers taking into account the effects of independent and non-identically distributed (i.n.i.d.) Nakagami-$m$ fading channels.

The common characteristic of the above mentioned works \cite{yilmaz2012novel,yilmaz2014computation,yilmaz2012computation,sagias2011higher}, however, is that they adopt the assumption of homogeneous fading channels. It has been proved that the homogeneous fading is often unrealistic since the surfaces are spatially correlated in practical propagation environments \cite{yacoub2007kappa}. Yet, very few results on the HOS of the channel capacity in non-homogeneous and composite fading conditions are available. Only recently, the HOS of the channel capacity for dispersed spectrum cognitive radio (CR) systems over i.n.i.d. $\eta$-$\mu$ fading channels was obtained in \cite{tsiftsis2015higher}. While these prior works have significantly improved our knowledge on the HOS of the channel capacity, a general analytic framework of spectrum aggregation systems which will account for more realistic fading models seems to be missing from the open literature. To address such non-homogeneous and composite fading environments, the generalized $\kappa$-$\mu$ and $\kappa$-$\mu$ shadowed fading channels are recently introduced in \cite{yacoub2007kappa,paris2014statistical,cotton2015human}, respectively. Compared with classic homogeneous fading models, the $\kappa$-$\mu$ and $\kappa$-$\mu$ shadowed fading models exhibit excellent agreement with measured land-mobile satellite, underwater acoustic, and body communications fading channels \cite{cotton2015human}. Moreover, the $\kappa$-$\mu$ fading channel includes the Rayleigh, Rician, and Nakagami-$m$ fading channels as special cases by setting the parameters $\kappa$ and $\mu$ to specific real positive values \cite{zhang2015effective}, while the $\kappa$-$\mu$ shadowed fading channel includes One-side Gaussian, Rayleigh, Rician, Nakagami-$m$, Hoyt, $\kappa$-$\mu$, $\eta$-$\mu$, and Rician shadowed fading channels as special cases \cite{Zhang2014effective}.

{
On the other hand, recent wireless applications become increasingly complex and require more realistic channel models for performance evaluation purposes \cite{dohler2011phy,guan2012measurement,zhang2012performance}. Because of the fact that the adopted fading models can describe a plethora of realistic fading propagation scenarios, they can serve as useful tools to this end.
}
Motivated by these important observations, we herein analytically investigate the HOS of the channel capacity for spectrum aggregation systems over $\kappa$-$\mu$ and $\kappa$-$\mu$ shadowed fading channels.
In particular, the main contributions of this paper can be summarized as:
\begin{itemize}
\item We first derive exact analytical expressions for the HOS of the channel capacity for spectrum aggregation systems over i.i.d., i.n.i.d $\kappa$-$\mu$ and i.i.d., correlated $\kappa$-$\mu$ shadowed fading channels, respectively. In contrast to exsiting works on second order statistics, the analysis of the HOS is still limited. It is worth noting that although the statistical characteristics of general fading models are very complicated, our derived results can be readily evaluated and efficiently programmed in most standard software packages (e.g., MATLAB and MATHEMATICA).
\item Furthermore, the asymptotically high- and low-SNR expressions for the HOS of the channel capacity are also presented to get additional insights into the impact of system parameters, such as fading parameters and number of aggregating frequency bands. More importantly, some of asymptotic expressions are given in terms of simple elementary functions.
\item With the help of the HOS of the channel capacity, we also provided useful performance metrics in terms of ergodic capacity, amount of fading (AOF), amount of dispersion (AOD), skewness, and kurtosis. Moreover, numerical results are provided to verify our analysis. Note that the presented analysis is very meaningful for communication systems to aggregate best available frequency bands in future spectrum-limited wireless networks.
\end{itemize}

This paper is organized as follows. The spectrum aggregation system and generalized $\kappa$-$\mu$ and $\kappa$-$\mu$ shadowed fading models are introduced in Section \ref{se:system}. In Section \ref{se:HOS}, we present the derivation of the HOS of the channel capacity and other important metrics. In Section \ref{se:num}, numerical results are shown to verify our present results. Finally, Section \ref{se:con} concludes the paper and summarizes the key findings.

\section{System and Channel Model}\label{se:system}
\begin{figure}[t]
\centering
\includegraphics[scale=0.85]{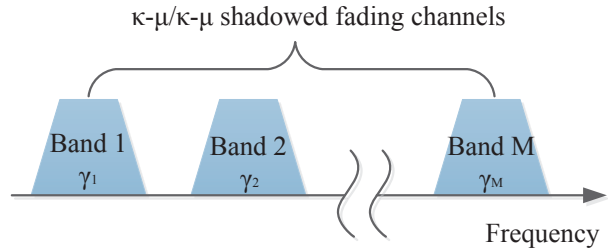}
\caption{Spectrum aggregation systems over generalized fading channels.
\label{fig:Spectrum_aggregation}}
\end{figure}

As illustrated in Fig. \ref{fig:Spectrum_aggregation}, the spectrum aggregation system exploits the benefits of frequency diversity by combining the instantaneous signal-to-noise ratios (SNRs), $\gamma_i$, from each noncontinuous band. By assuming $M$ available frequency diversity bands, the end-to-end SNR, $\gamma$, at the output of each UE's receiver is given by $\gamma = \sum\nolimits_{i = 1}^M {{\gamma _i}} $.
Moreover, each frequency diversity channel is assumed to be slow and frequency non-selective. Note that the end-to-end SNR $\gamma$ has a similar form of the SNR at the output of an MRC combiner.

\vspace{-2mm}
\subsection{$\kappa$-$\mu$ fading channels}
The $\kappa$-$\mu$ distribution can be regarded as a generalization of the classic Rician fading model for line-of-sight (LoS) scenarios, and has been extensively used in spatially non-homogeneous propagation environments. The $\kappa$-$\mu$ fading signal is a composition of clusters of multipath waves with scattered waves of identical power with a dominant component of arbitrary power found within each cluster. Furthermore, the parameter $\kappa$ represents the ratio between the total power of the dominant components and the total power of the scattered waves, while $\mu$ is the number of clusters.
The probability density function (PDF) of the sum of $M$ i.i.d. squared $\kappa$-$\mu$ random variables (RVs) is given by \cite[Eq. (10)]{yacoub2007kappa}
\begin{align}
&{f_{i.i.d}}\left( \gamma  \right) = \frac{{\mu M{{\left( {1 + \kappa } \right)}^{\frac{{\mu M + 1}}{2}}}{\gamma ^{\frac{{\mu M - 1}}{2}}}}}{{{e^{\mu M\kappa }}{\kappa ^{\frac{{\mu M - 1}}{2}}}{{\left( {\Omega M} \right)}^{\frac{{\mu M + 1}}{2}}}}}\exp \left( { - \frac{{\mu \left( {1 + \kappa } \right)\gamma }}{\Omega }} \right)\notag \\
&\times {I_{\mu M - 1}}\left( {2\mu \sqrt {\frac{{\kappa \left( {1 + \kappa } \right)M\gamma }}{\Omega }} } \right)\label{eq:iid_kappa_mu_pdf_1}\\
&= \frac{{\mu M{{\left( {1 + \kappa } \right)}^{\frac{{\mu M + 1}}{2}}}{\gamma ^{\frac{{\mu M - 1}}{2}}}}}{{{e^{\mu M\kappa }}{\kappa ^{\frac{{\mu M - 1}}{2}}}{{\left( {\Omega M} \right)}^{\frac{{\mu M + 1}}{2}}}}}\exp \left( { - \frac{{\mu \left( {1 + \kappa } \right)\gamma }}{\Omega }} \right)\notag \\
&\times\sum\limits_{i = 0}^\infty  {\frac{1}{{i!\Gamma \left( {\mu M + i} \right)}}{{\left( {\mu \sqrt {\frac{{\kappa \left( {1 + \kappa } \right)M\gamma }}{\Omega }} } \right)}^{\mu M + 2i - 1}}}, \label{eq:iid_kappa_mu_pdf}
\end{align}
where $\Omega$ denotes the average SNR of each $\kappa$-$\mu$ RV, $I_v(\cdot)$ is the modified Bessel function of first kind \cite[Eq. (8.406.1)]{gradshtein2000table}, and $\Gamma(\cdot)$ denotes the Gamma function \cite[Eq. (8.31)]{gradshtein2000table}. From \eqref{eq:iid_kappa_mu_pdf_1} to \eqref{eq:iid_kappa_mu_pdf}, we have used the identity of \cite[Eq. (8.445)]{gradshtein2000table} and carried out some algebraic manipulations.

The PDF of the sum of $M$ i.n.i.d. squared $\kappa$-$\mu$ RVs is given by \cite[Eq. (4)]{peppas2012sum}
\begin{align}
{f_{i.n.i.d}}\left( \gamma  \right) &= \frac{{{e^{ - \frac{\gamma }{{2\beta }}}}{\gamma ^{U - 1}}}}{{{{\left( {2\beta } \right)}^U}\Gamma \left( U \right)}}\sum\limits_{k = 0}^\infty  {\frac{{k!{c_k}}}{{{{\left( U \right)}_k}}}L_k^{\left( {U - 1} \right)}\left( {\frac{{U\gamma }}{{2\beta \xi }}} \right)} \notag \\
& =\frac{{{e^{ - \frac{\gamma }{{2\beta }}}}}}{{{{\left( {2\beta } \right)}^U}}}\sum\limits_{k = 0}^\infty  {{c_k}\sum\limits_{q = 0}^k {\frac{{{{\left( { - k} \right)}_q}{\gamma ^{q + U- 1}}}}{{q!\Gamma \left( {U+ q} \right)}}{{\left( {\frac{U}{{2\beta \xi }}} \right)}^q}} },\label{eq:inid_kappa_mu_pdf}
\end{align}
where $(a)_b$ denotes the Pochhammer symbol \cite{gradshtein2000table}, $U = \sum\nolimits_{i = 1}^M {{\mu _i}}$, and the series representation of generalized Laguerre polynomial $L_k^v\left( \cdot \right)$ has been used as \cite[Eq. (05.08.02.0001.01)]{Wolfram2011function}
\begin{align}
L_k^v\left( y \right) = \frac{{\Gamma \left( {v + k + 1} \right)}}{{k!}}\sum\limits_{q = 0}^k
 {\frac{{{{\left( { - k} \right)}_q}{y^q}}}{{q!\Gamma \left( {v + q + 1} \right)}}}.
\end{align}
{Moreover, the coefficients $c_k$ in \eqref{eq:inid_kappa_mu_pdf} can be obtained as}
\begin{align}\label{eq:c_k}
{c_k} = \frac{1}{k}\sum\limits_{j = 0}^{k - 1} {{c_j}{d_{k - j}}}, \;\;\; k \geqslant 1
\end{align}
\begin{align}\label{eq:c_0}
{d_j} &\triangleq  - \frac{{j\beta U}}{{2\xi }}{\sum\limits_{i = 1}^M {{\chi _i}{a_i}{{\left( {\beta  - {a_i}} \right)}^{j - 1}}\left( {\frac{\xi }{{\beta \xi  + {a_i}\left( {U - \xi } \right)}}} \right)} ^{j + 1}} \notag \\
&+ {\sum\limits_{i = 1}^M {{\mu _i}\left( {\frac{{1 - {a_i}/\beta }}{{1 + \left( {{a_i}/\beta } \right)\left( {U/\xi  - 1} \right)}}} \right)} ^j}, \;\;\; j \geqslant 1  \\
{c_0} &\triangleq {\left( {\frac{U}{\xi }} \right)^U}\exp \left( { - \frac{1}{2}\sum\limits_{i = 1}^M {\frac{{{\chi _i}{a_i}\left( {U - \xi } \right)}}{{\beta \xi  + {a_m}\left( {U - \xi } \right)}}} } \right)\notag \\
&\times {\prod\limits_{i = 1}^M {\left( {1 + \frac{{{a_i}}}{\beta }\left( {U/\xi  - 1} \right)} \right)} ^{ - {\mu _i}}},
\end{align}
where ${\chi _i} \triangleq 2{\mu _i}{\kappa _i}$ and ${a_i} \triangleq {\Omega _i}/2{\mu _i}\left( {1 + {\kappa _i}} \right)$. In order to guarantee the uniform convergence of \eqref{eq:inid_kappa_mu_pdf}, the parameters $\xi$ and $\beta$ should be chosen appropriately \cite{castano2005distribution}.

\subsection{$\kappa$-$\mu$ shadowed fading channels}
Similar to the the same multipath/shadowing scheme used in the Rician shadowed model, a natural generalization of the $\kappa$-$\mu$ distribution can be obtained by an LoS shadow fading model. Unlike the $\kappa$-$\mu$ distribution, the $\kappa$-$\mu$ shadowed model assumes that all the dominant components are subject to the same common fluctuation due to shadowing. With the assumption of shadowing components are correlated, while multipath components are uncorrelated, the PDF of the sum of $M$ i.i.d. squared $\kappa$-$\mu$ shadowed RVs is given by \cite{bhatnagar2015sum}
\begin{align}\label{eq:iid_kappa_mu_shadowed_pdf_1}
&{f_{i.i.d.}}\left( \gamma  \right)= {\left( {\frac{{\mu M\left( {1 + \kappa } \right)}}{{\bar \gamma }}} \right)^{\mu M}}{\left( {\frac{m}{{m + \kappa \mu }}} \right)^{mM}}\frac{{{\gamma ^{\mu M - 1}}}}{{\Gamma \left( {\mu M} \right)}}\notag \\
&\times {e^{ - \frac{{\mu M\left( {1 + \kappa } \right)\gamma }}{{\bar \gamma }}}}{}_1{F_1}\left( {mM,\mu M;\frac{{M{\mu ^2}\kappa \left( {1 + \kappa } \right)\gamma }}{{\left( {\mu \kappa  + m} \right)\bar \gamma }}} \right)
\end{align}
where $m$ denotes the shaping parameter of shadowing, and ${}_1{F_1}\left(\cdot\right)$ is the confluent hypergeometric function \cite[Eq. (9.210.1)]{gradshtein2000table}. By utilizing the following identity
\begin{align}\label{eq:1F1}
{}_1{F_1}\left( {a,b;x} \right) = \sum\limits_{q = 0}^\infty  {\frac{{{{\left( a \right)}_q}{x^q}}}{{{{\left( b \right)}_q}q!}}},
\end{align}
we can rewrite \eqref{eq:iid_kappa_mu_shadowed_pdf_1} in an alternative form as
\begin{align}\label{eq:iid_kappa_mu_shadowed_pdf}
{f_{i.i.d.}}\left( \gamma  \right) &= {\left( {\frac{{\mu M\left( {1 + \kappa } \right)}}{{\bar \gamma }}} \right)^{\mu M}}{\left( {\frac{m}{{m + \kappa \mu }}} \right)^{mM}}\frac{{{e^{ - \frac{{\mu M\left( {1 + \kappa } \right)\gamma }}{{\bar \gamma }}}}}}{{\Gamma \left( {\mu M} \right)}}\notag \\
&\times \sum\limits_{q = 0}^\infty  {\frac{{{{\left( {mM} \right)}_q}{{\left( {\frac{{M{\mu ^2}\kappa \left( {1 + \kappa } \right)}}{{\left( {\mu \kappa  + m} \right)\bar \gamma }}} \right)}^q}{\gamma ^{\mu M + q - 1}}}}{{{{\left( {\mu M} \right)}_q}q!}}} ,
\end{align}
where only elementary functions appear. Thus, \eqref{eq:iid_kappa_mu_shadowed_pdf} can facilitate the calculation involved the PDF expression of $\kappa$-$\mu$ shadowed fading channels.

Furthermore, the PDF of the sum of $M$ correlated squared $\kappa$-$\mu$ shadowed RVs is given by \cite[Eq. (16)]{bhatnagar2015sum}
\begin{align}\label{eq:cor_kappa_mu_shadowed_pdf_1}
{f_{cor}}\left( \gamma  \right) &= A{\left( {\frac{\eta }{{\bar \gamma }}} \right)^U}{\gamma ^{U - 1}}{e^{ - \frac{\eta \gamma}{{\bar \gamma }}}}\notag \\
&\times \sum\limits_{k = 0}^\infty  {{D_k}} {}_1{F_1}\left( {mM+k,U;\frac{{\eta \gamma }}{{\bar \gamma \left( {1 + {\lambda ^{ - 1}}} \right)}}} \right),
\end{align}
where ${\bar \gamma }$ denotes the average SNR, $A \triangleq {\prod\nolimits_{i = 1}^M {\left( {\frac{\lambda }{{{\lambda _i}}}} \right)} ^m}$, $\eta \triangleq \sum\nolimits_{i = 1}^M {{\mu _i}} \left( {1 + {\kappa _i}} \right)$, and
\begin{align}\label{eq:Dk}
{D_k} &= \frac{{{\delta _k}}}{{{\lambda ^{mM + k}}\Gamma \left( U \right)}}{\left( {1 + {\lambda ^{ - 1}}} \right)^{ - \left( {mM + k} \right)}},\\
{\delta _{k+1}} &=\frac{m}{{k + 1}}\sum\limits_{q = 1}^{k + 1} {\sum\limits_{i = 1}^M {{{\left( {1 - \frac{\lambda }{{{\lambda _i}}}} \right)}^q}} } {\delta _{k + 1 - i}},\\
{\delta _0} &= 1.
\end{align}
Moreover, $\lambda  \triangleq \min \left( {{\lambda _1},{\lambda _2}, \cdots ,{\lambda _M}} \right)$ is the minimum eigenvalue of the matrix $\mathbf{DC}$ with ${\bf{D}} = {\tt diag}\left\{ {\frac{{{\mu _i}{\kappa _i}}}{m}} \right\}$ represents a diagonal matrix and ${\bf{C}}$ denotes the $M \times M$ positive definite matrix given by
\begin{align}\label{eq:C}
{\bf{C}} \triangleq \left[ {\begin{array}{*{20}{c}}
1&{\sqrt {{\rho _{12}}} }& \cdots &{\sqrt {{\rho _{1M}}} }\\
{\sqrt {{\rho _{21}}} }&1& \cdots &{\sqrt {{\rho _{2M}}} }\\
 \vdots & \vdots & \ddots & \vdots \\
{\sqrt {{\rho _{M1}}} }& \cdots & \cdots &1
\end{array}} \right],
\end{align}
where $0 \leq{{\rho _{pq}}}\leq1, 1\leq p,q \leq M$ denotes the correlation coefficient of the dominating components of $\kappa$-$\mu$ shadowed RVs.
With the help of \eqref{eq:1F1}, we can rewrite \eqref{eq:cor_kappa_mu_shadowed_pdf_1} as
\begin{align}\label{eq:cor_kappa_mu_shadowed_pdf}
{f_{cor}}\left( \gamma  \right) &= A{\left( {\frac{\eta }{{\bar \gamma }}} \right)^U}{\gamma ^{U - 1}}{e^{ - \frac{\eta \gamma}{{\bar \gamma }}}}\sum\limits_{k = 0}^\infty  {{D_k}} \notag \\
&\times{\sum\limits_{q = 0}^\infty  {\frac{{{{\left( {mM+k} \right)}_q}}}{{{{\left( U \right)}_q}q!}}\left( {\frac{{\eta \gamma }}{{\bar \gamma \left( {1 + {\lambda ^{ - 1}}} \right)}}} \right)} ^q},
\end{align}

\section{Higher-Order Capacity Statistics}\label{se:HOS}
In this section, we present the statistical analysis for the derivation of the HOS of the channel capacity for spectrum aggregation systems. Without loss of generality, the HOS of the channel capacity can be defined as \cite{yilmaz2012computation,tsiftsis2015higher}
\begin{align}\label{eq:HOS}
{\Lambda _n} = {\tt E}\left( {\log _2^n\left( {1 + \gamma } \right)} \right),
\end{align}
where $n \in \mathbb{N}$ is the order of the capacity statistics, and ${\tt E} \left( \cdot \right)$ denotes the expectation operator. Note that the first-order statistics of channel capacity is well-known as the ergodic capacity. Without loss of generality, the HOS of the channel capacity for the spectrum aggregation systems with $M$ non-adjacent frequency bands is given by
\begin{align}\label{eq:HOS_integral}
{\Lambda _n} &= \int_0^\infty  \int_0^\infty   \cdots \int_0^\infty  {{\log }^n}\left( {1 + \sum\limits_{i = 1}^M {{\gamma _i}} } \right)\notag \\
&\times {f_\gamma }\left( {{\gamma _1},{\gamma _2}, \cdots {\gamma _M}} \right)d{\gamma _1}d{\gamma _2} \cdots d{\gamma _M} ,
\end{align}
where ${f_\gamma }\left( {{\gamma _1},{\gamma _2}, \cdots {\gamma _M}} \right) $ represents the joint pdf of the instantaneous SNRs of each band. Unfortunately, it is very tedious and computationally cumbersome to obtain the joint pdf even for the simple i.i.d. case. One possible way to solve this problem is to use the moment generating function (MGF) based method proposed in \cite{yilmaz2012computation}. However, the HOS of the channel capacity is given in terms of a single-integral expression, which makes it difficult to be mathematically employed. Therefore, in the following, we derive the HOS of the channel capacity for spectrum aggregation systems over generalized fading channels by utilizing the pdf of the total SNR $\gamma$. It is worthy to mention that all our derived results are given in analytical form, which is easy to show the key impacts of system performance.

\subsection{$\kappa$-$\mu$ fading channels}
We first consider the higher-order capacity statistics of spectrum aggregation systems over i.i.d. $\kappa$-$\mu$ fading channels as follows.
\begin{thm}\label{th:iid_kappa_mu_HOS}
The higher-order capacity statistics of spectrum aggregation systems over i.i.d. $\kappa$-$\mu$ fading channels can be expressed as
\begin{align}\label{eq:iid_kappa_mu_HOS}
{\Lambda _n} &=  \frac{1}{{{e^{\mu M\kappa }} {{\ln }^n}2}}\sum\limits_{i = 0}^\infty  \frac{{{{\left( {\mu \kappa M} \right)}^i}}}{{i!\Gamma \left( {\mu M + i} \right)}}{{\left( {\frac{{\mu \left( {1 + \kappa } \right)}}{\Omega }} \right)}^{\mu M + i}}\notag \\
&\times J\left( {\mu M + i,\frac{{\mu \left( {1 + \kappa } \right)}}{\Omega },n} \right)  ,
\end{align}
where the auxiliary function $J(\cdot)$ is given by \eqref{eq:J_result} in the Appendix A.
\end{thm}
\begin{IEEEproof}
Substituting \eqref{eq:iid_kappa_mu_pdf} into \eqref{eq:HOS}, we can derive
\begin{align}\label{eq:iid_kappa_mu_HOS_1}
{\Lambda _n} &= \frac{{\mu M{{\left( {1 + \kappa } \right)}^{\frac{{\mu M + 1}}{2}}}}}{{{e^{\mu M\kappa }}{\kappa ^{\frac{{\mu M - 1}}{2}}}{{\left( {\Omega M} \right)}^{\frac{{\mu M + 1}}{2}}{\ln ^n}{2}}}}\notag \\
&\times \sum\limits_{i = 0}^\infty  \frac{1}{{i!\Gamma \left( {\mu M + i} \right)}}{{\left( {\mu \sqrt {\frac{{\kappa \left( {1 + \kappa } \right)M}}{\Omega }} } \right)}^{\mu M + 2i - 1}}\notag \\
&\times \int_0^\infty  {{{\ln }^n}\left( {1 + \gamma } \right){\gamma ^{\mu M + i - 1}}{e^{ - \frac{{\mu \left( {1 + \kappa } \right)}}{\Omega }\gamma}}d\gamma}  .
\end{align}
{To the best of the authors' knowledge, the integral in \eqref{eq:iid_kappa_mu_HOS_1} is not included in tables of classical reference books such as \cite{gradshtein2000table}.
Nervertheless, as shown in Appendix A, it can be computed in closed form thus completing the proof.}
\end{IEEEproof}

{Note that the auxiliary function $J(\cdot)$ requires $\mu$ is restricted to integer values, which assumes finite numbers of multipath clusters. In the most general case of real $\mu$, integrals of the form
\begin{equation}\label{eq:Intk}
\mathcal{K}(\nu, \mu, a) = \int_0^{\infty} \ln^\nu(1+x)\exp(-a x) x^{\mu} \mathrm{d}x
\end{equation}
should be evaluated. By performing the change of variables $a x = y^2$, \eqref{eq:Intk} can be expressed as
\begin{equation}\label{eq:Intk2}
\mathcal{K}(\nu, \mu, a) = \frac{2}{a^{\mu+1}}\int_0^{\infty} \ln^\nu\left(1+\frac{y^2}{a}\right)\exp(-y^2) y^{2\mu+1} \mathrm{d}y.
\end{equation}
This integral can be evaluated numerically in an efficient manner by employing a $N$-point Gauss-Chebyshev quadrature rule as
\begin{equation}\label{eq:Intk3}
\mathcal{K}(\nu, \mu, a) = \frac{2}{a^{\mu+1}} \sum_{k=1}^{15} w_k \ln^\nu\left(1+\frac{t_k^2}{a}\right) t_k^{2\mu+1}
\end{equation}
where $w_k$ and $t_k$ are the weights and abscissae given in \cite{steen1969gaussian}. Note that we only need 15 terms in \eqref{eq:Intk3} to converge adequately.
}

By taking $n=1$ in \eqref{eq:iid_kappa_mu_HOS}, we can obtain the first-order statistics of the channel capacity, which is the well-known ergodic capacity as
\begin{align}\label{eq:iid_kappa_mu_capacity}
{\Lambda _1} &=  \frac{{{e^{\frac{{\mu \left( {1 + \kappa } \right)}}{\Omega } - \mu M\kappa }}}}{{{{\ln }^n}2}}\sum\limits_{i = 0}^\infty  \frac{{{{\left( {\mu \kappa M} \right)}^i}}}{{i!\Gamma \left( {\mu M + i} \right)}} \notag \\
&\times \sum\limits_{k = 0}^{\mu M + i - 1}  \Bigg[ {{\left( { - { {\frac{{\mu \left( {1 + \kappa } \right)}}{\Omega }}  }} \right)}^{\mu M + i - k - 1}}\left( {\begin{array}{*{20}{c}}
{\mu M + i - 1}\\
k
\end{array}} \right)\notag \\
&\times G_{2,3}^{3,0}\left( {\frac{{\mu \left( {1 + \kappa } \right)}}{\Omega }\left| {\begin{array}{*{20}{c}}
{1,1}\\
{0,0,1 + k}
\end{array}} \right.} \right) \Bigg],
\end{align}
where $\left( {\begin{array}{*{10}{c}}
a\\
b
\end{array}} \right) = \frac{a!}{b!(a-b)!}$, and $G(\cdot)$ denotes the Meijer's $G$-function \cite[Eq. (9.301)]{gradshtein2000table}.

\begin{lemm}\label{lemm:high_iid_kappa_mu_HOS_high}
For the high- and low-SNR regimes, the higher-order capacity statistics of spectrum aggregation systems over i.i.d. $\kappa$-$\mu$ fading channels can be respectively expressed as
\begin{align}
\Lambda_n^\infty  &= \frac{1}{{{e^{\mu M\kappa }}{{\ln }^n}2}}\sum\limits_{i = 0}^\infty  \frac{{{{\left( {\mu \kappa M} \right)}^i}}}{{i!\Gamma \left( {\mu M + i} \right)}}{{\left( {\frac{{\mu \left( {1 + \kappa } \right)}}{\Omega }} \right)}^{\mu M + i}} \notag \\
&\times Q\left( {\mu M + i-1,\frac{{\mu \left( {1 + \kappa } \right)}}{\Omega },n} \right) ,\label{eq:high_iid_kappa_mu_HOS}\\
\Lambda _n^{\gamma  \to 0} &= \frac{{n!}}{{{e^{\mu M\kappa }}{{\ln }^n}2}}\sum\limits_{k = 0}^\infty  \frac{{S_{k + n}^n}}{{\left( {k + n} \right)!}}\frac{{\Gamma \left( {k + n + \mu M} \right)}}{{\Gamma \left( {\mu M} \right)}}\notag \\
&\times {{\left( {\frac{\Omega }{{\mu \left( {1 + \kappa } \right)}}} \right)}^{n + k}}{}_1{F_1}\left( {k + n + \mu M;\mu M;\mu \kappa M} \right) ,\label{eq:low_iid_kappa_mu_HOS}
\end{align}
where $S_m^n$ is the Stirling number of the first kind \cite[Eq. (9.740)]{gradshtein2000table}, and the auxiliary function $Q(\cdot)$ is given by \eqref{eq:Q_result} in the Appendix. B.
\end{lemm}
\begin{IEEEproof}
By taking large values of $\gamma$ in \eqref{eq:HOS} and using \eqref{eq:iid_kappa_mu_pdf}, the higher-order capacity is given by
\begin{align}\label{eq:high_iid_kappa_mu_HOS_1}
\Lambda _n^\infty &= \frac{{\mu M{{\left( {1 + \kappa } \right)}^{\frac{{\mu M + 1}}{2}}}}}{{{e^{\mu M\kappa }}{\kappa ^{\frac{{\mu M - 1}}{2}}}{{\left( {\Omega M} \right)}^{\frac{{\mu M + 1}}{2}}}}{\ln ^n}{2}}\sum\limits_{i = 0}^\infty  \frac{1}{{i!\Gamma \left( {\mu M + i} \right)}}\notag \\
&\times {{\left( {\mu \sqrt {\frac{{\kappa \left( {1 + \kappa } \right)M}}{\Omega }} } \right)}^{\mu M + 2i - 1}}\notag \\
&\times \int_0^\infty  {{{\ln }^n}\left( {  \gamma } \right){\gamma ^{\mu M + i - 1}}{e^{ - \frac{{\mu \left( {1+ \kappa } \right)}}{\Omega }\gamma}}d\gamma}.
\end{align}
With the aid of \cite[Eq. (2.5.1.7)]{prudnikov1990integrals3}, the integral in \eqref{eq:high_iid_kappa_mu_HOS_1} can be calculated as
\begin{align}\label{eq:high_iid_kappa_mu_HOS_2}
\Lambda _n^\infty  &= \frac{{\mu M{{\left( {1 + \kappa } \right)}^{\frac{{\mu M + 1}}{2}}}}}{{{e^{\mu M\kappa }}{\kappa ^{\frac{{\mu M - 1}}{2}}}{{\left( {\Omega M} \right)}^{\frac{{\mu M + 1}}{2}}}{{\ln }^n}2}}\notag \\
&\times \sum\limits_{i = 0}^\infty  \frac{1}{{i!\Gamma \left( {\mu M + i} \right)}}{{\left( {\mu \sqrt {\frac{{\kappa \left( {1 + \kappa } \right)M}}{\Omega }} } \right)}^{\mu M + 2i - 1}}\notag \\
&\times \frac{{{d^n}}}{{d{a^n}}}{{\left( {\frac{{\Gamma \left( {a + 1} \right)}}{{{{\left( {\mu \left( {1 + \kappa } \right)/\Omega } \right)}^{a + 1}}}}} \right)}_{a = \mu M + i-1}}.
\end{align}
Then, the high-SNR HOS \eqref{eq:high_iid_kappa_mu_HOS} can be derived by using \eqref{eq:Q_result} and after some algebraic manipulation.

Moreover, the low-SNR HOS can be obtained by taking $\rho \to 0$, and using the well-known expansion of the logarithm function as \cite[Eq. (9.741.2)]{gradshtein2000table}
\begin{align}\label{eq:log_expansion}
{\ln ^n}\left( {1 + z} \right) = n!\sum\limits_{k = 0}^\infty  {S_{k + n}^n\frac{{{z^{k + n}}}}{{\left( {k + n} \right)!}}}, \;\;\;z \to 0
\end{align}
Substituting \eqref{eq:log_expansion} into \eqref{eq:HOS}, and with the aid of \eqref{eq:iid_kappa_mu_pdf} and \eqref{eq:HOS}, we derive the low-SNR HOS as
\begin{align}\label{eq:low_HOS_integral}
\Lambda _n^{\gamma  \to 0} &= \frac{{\mu M{{\left( {1 + \kappa } \right)}^{\frac{{\mu M + 1}}{2}}}n!}}{{{e^{\mu M\kappa }}{\kappa ^{\frac{{\mu M - 1}}{2}}}{{\left( {\Omega M} \right)}^{\frac{{\mu M + 1}}{2}}}{{\ln }^n}2}}\sum\limits_{k = 0}^\infty  {\frac{{S_{k + n}^n}}{{\left( {k + n} \right)!}}} \notag \\
&\times \int_{\rm{0}}^\infty  {\gamma ^{k + n + \frac{{\mu M - 1}}{2}}}\exp \left( { - \frac{{\mu \left( {1 + \kappa } \right)\gamma }}{\Omega }} \right)\notag \\
&\times {I_{\mu M - 1}}\left( {2\mu \sqrt {\frac{{\kappa \left( {1 + \kappa } \right)M\gamma }}{\Omega }} } \right)d\gamma .
\end{align}
To evaluate the integral in \eqref{eq:low_HOS_integral}, we can utilize the following identity \cite[Eq. (3.15.2.5)]{prudnikov1992integrals4}
\begin{align}\label{eq:low_HOS_integral_identity}
&\int_{\rm{0}}^\infty  {{x^q}\exp \left( { - px} \right){I_v}\left( {a\sqrt x } \right)dx}  =\notag \\
& \frac{{\Gamma \left( {q \!+\! v/2 \!+\! 1} \right)}}{{\Gamma \left( {v + 1} \right)}}\frac{{{{\left( {a/2} \right)}^v}}}{{{p^{q \!+\! v/2 \!+\! 1}}}}{}_1{F_1}\left( {q \!+\! \frac{v}{2} \!+\! 1;v \!+\! 1;\frac{{{a^2}}}{{4p}}} \right).
\end{align}
Finally, we arrive at the desired result in \eqref{eq:low_iid_kappa_mu_HOS} after some basic algebra.
\end{IEEEproof}

{Note that the auxiliary function $Q(\cdot)$ can apply for arbitrary positive real values of $\mu$, so the asymptotical results are generalized.}
It is easy to see from \eqref{eq:high_iid_kappa_mu_HOS} and \eqref{eq:low_iid_kappa_mu_HOS} that the HOS of the channel capacity is an increasing function in the average SNR $\Omega$ and $M$.

\begin{thm}
The higher-order capacity statistics of spectrum aggregation systems over i.n.i.d. $\kappa$-$\mu$ fading channels can be expressed as
\begin{align}\label{eq:inid_kappa_mu_HOS}
{\Lambda _n} &= \frac{1}{{{{\left( {2\beta } \right)}^U}}{{{\ln }^n}2}}\sum\limits_{k = 0}^\infty  {{c_k}\sum\limits_{q = 0}^k {\frac{{{{\left( { - k} \right)}_q} }}{{q!\Gamma \left( {U + q} \right)}}{{\left( {\frac{U}{{2\beta \xi }}} \right)}^q}} } \notag \\
&\times J\left( {q + U,\frac{1}{{2\beta }},n} \right).
\end{align}
\end{thm}
\begin{IEEEproof}
The proof is readily completed by taking \eqref{eq:inid_kappa_mu_pdf} into \eqref{eq:HOS}, and using \eqref{eq:J_result}.
\end{IEEEproof}

\begin{lemm}\label{lemm:high_inid_kappa_mu_HOS_high}
For the high- and low-SNR regime, the higher-order capacity statistics of spectrum aggregation systems over i.n.i.d. $\kappa$-$\mu$ fading channels can be expressed as
\begin{align}
\Lambda _n^\infty &= \frac{1}{{{{\left( {2\beta } \right)}^U}}{{{\ln }^n}2}}\sum\limits_{k = 0}^\infty  {{c_k}\sum\limits_{q = 0}^k {\frac{{{{\left( { - k} \right)}_q} }}{{q!\Gamma \left( {U + q} \right)}}{{\left( {\frac{U}{{2\beta \xi }}} \right)}^q}} }\notag \\
&\times Q\left( {q +U-1,\frac{1}{{2\beta }},n} \right),\label{eq:high_inid_kappa_mu_HOS}\\
\Lambda _n^{\gamma  \to 0} &= \frac{{n!}}{{{{\ln }^n}2}}\sum\limits_{k = 0}^\infty  {c_k}\sum\limits_{q = 0}^k \frac{{{{\left( { - k} \right)}_q}}}{{q!\Gamma \left( {U + q} \right)}}{{\left( {\frac{U}{\xi }} \right)}^q}\notag \\
&\times \sum\limits_{p = 0}^\infty  {\frac{{S_{p + n}^n\Gamma \left( {U + q + p + n} \right)}}{{\left( {p + n} \right)!{{\left( {2\beta } \right)}^{ - \left( {p + n} \right)}}}}}  .\label{eq:low_inid_kappa_mu_HOS}
\end{align}
\end{lemm}
\begin{IEEEproof}
With the help of \cite[Eq. (3.351.3)]{gradshtein2000table}, the proof can be completed by following similar steps in Lemma \ref{lemm:high_iid_kappa_mu_HOS_high}.
\end{IEEEproof}
Note that the low-SNR HOS of the channel capacity \eqref{eq:low_inid_kappa_mu_HOS} is given in terms of simple elementary functions. Therefore, we can obtain the implication that the HOS is an increasing function in $U$.

\subsection{$\kappa$-$\mu$ shadowed fading channels}
Now, we move on to consider the higher-order capacity statistics for $\kappa$-$\mu$ shadowed fading channels. As a first step, the case of i.i.d. $\kappa$-$\mu$ shadowed fading channel is investigated.
\begin{thm}\label{th:iid_kappa_mu_shadowed_HOS}
The higher-order capacity statistics of spectrum aggregation systems over i.i.d. $\kappa$-$\mu$ shadowed fading channels can be expressed as
\begin{align}\label{eq:iid_kappa_mu_shadowed_HOS}
{\Lambda _n} &= {\left( {\frac{{\mu M\left( {1 + \kappa } \right)}}{{\bar \gamma }}} \right)^{\mu M}}{\left( {\frac{m}{{m + \kappa \mu }}} \right)^{mM}}\frac{{1}}{{\Gamma \left( {\mu M} \right)}{{{\ln }^n}2}}\notag \\
&\times \sum\limits_{q = 0}^\infty  \frac{{{{\left( {mM} \right)}_q}{{\left( {\frac{{M{\mu ^2}\kappa \left( {1 + \kappa } \right)}}{{\left( {\mu \kappa  + m} \right)\bar \gamma }}} \right)}^q}}}{{{{\left( {\mu M} \right)}_q}q!}}\notag \\
&\times J\left( {q + \mu M,\frac{{\mu M\left( {1 + \kappa } \right)}}{{\bar \gamma }},n} \right) .
\end{align}
\end{thm}
\begin{IEEEproof}
This result is a direct consequence of taking \eqref{eq:iid_kappa_mu_shadowed_pdf} into \eqref{eq:HOS}, and using \eqref{eq:J_result}.
\end{IEEEproof}

\begin{lemm}\label{lemm:high_iid_kappa_mu_shadowed_HOS}
For the high- and low-SNR regimes, the higher-order capacity statistics of spectrum aggregation systems over i.i.d. $\kappa$-$\mu$ shadowed fading channels can be expressed as
\begin{align}
\Lambda _n^\infty &= {\left( {\frac{{\mu M\left( {1 + \kappa } \right)}}{{\bar \gamma }}} \right)^{\mu M}}{\left( {\frac{m}{{m + \kappa \mu }}} \right)^{mM}}\frac{{1}}{{\Gamma \left( {\mu M} \right)}{{{\ln }^n}2}}\notag \\
&\times \sum\limits_{q = 0}^\infty  \frac{{{{\left( {mM} \right)}_q}{{\left( {\frac{{M{\mu ^2}\kappa \left( {1 + \kappa } \right)}}{{\left( {\mu \kappa  + m} \right)\bar \gamma }}} \right)}^q}}}{{{{\left( {\mu M} \right)}_q}q!}}\notag \\
&\times Q\left( {q + \mu M-1,\frac{{\mu M\left( {1 + \kappa } \right)}}{{\bar \gamma }},n} \right),\label{eq:high_iid_kappa_mu_shadowed_HOS}\\
\Lambda _n^{\gamma  \to 0} &= {\left( {\frac{m}{{m + \kappa \mu }}} \right)^{mM}}\frac{{n!}}{{\Gamma \left( {\mu M} \right)}}\notag \\
&\sum\limits_{k = 0}^\infty  {\frac{{S_{k + n}^n}}{{\left( {k + n} \right)!}}} \frac{{\Gamma \left( {k + n + \mu M} \right)}}{{{{\left( {\mu M\left( {1 + \kappa } \right)/\bar \gamma } \right)}^{k + n}}}}\notag \\
&\times {}_2{F_1}\left( {mM,k + n + \mu M;\mu M;\frac{{\mu \kappa }}{{\mu \kappa  + m}}} \right).\label{eq:low_iid_kappa_mu_shadowed_HOS}
\end{align}
\end{lemm}
\begin{IEEEproof}
We can obtain \eqref{eq:high_iid_kappa_mu_shadowed_HOS} by following similar steps in Theorem \ref{th:iid_kappa_mu_shadowed_HOS}. While for the case of the low-SNR regime, the integral identity \cite[Eq. (3.35.1.2)]{prudnikov1992integrals4}
\begin{align}\label{eq:integral_1F1}
\int_0^\infty  {{x^q}{e^{ - px}}} {}_1{F_1}\left( {a,b;\omega x} \right)dx = \frac{{\Gamma \left( {q \!+\! 1} \right)}}{{{p^{q \!+\! 1}}}}{}_2{F_1}\left( {a,q \!+\! 1;b;\frac{\omega }{p}} \right)
\end{align}
should be invoked. Note that condition on the arguments of \eqref{eq:integral_1F1}, $p > 0$, $p > \omega$, and $q > -1$, is satisfied.
\end{IEEEproof}

Considering the case where the frequency bands are correlated, we present the following theorem.
\begin{thm}
The higher-order capacity statistics of spectrum aggregation systems over correlated $\kappa$-$\mu$ shadowed fading channels can be expressed as
\begin{align}\label{eq:cor_kappa_mu_shadowed_HOS}
{\Lambda _n} &= \frac{A}{{{\ln }^n}2}{\left( {\frac{\eta }{{\bar \gamma }}} \right)^U}\sum\limits_{k = 0}^\infty  {{D_k}} {\sum\limits_{q = 0}^\infty  {\frac{{{{\left( {mM+k} \right)}_q}}}{{{{\left( U \right)}_q}q!}}\left( {\frac{\eta }{{\bar \gamma \left( {1 + {\lambda ^{ - 1}}} \right)}}} \right)} ^q}\notag \\
&\times J\left( {q + U,\frac{\eta }{{\bar \gamma }},n} \right) .
\end{align}
\end{thm}
\begin{IEEEproof}
The proof can be completed by following similar steps in Theorem \ref{th:iid_kappa_mu_HOS}.
\end{IEEEproof}
We note from \eqref{eq:cor_kappa_mu_shadowed_HOS} that the HOS of the channel capacity is an increasing function in the number of frequency bands $M$ and shadowing parameter $m$ and as such it obtains its maximum value for $m \to \infty$.

\begin{lemm}\label{lemm:high_cor_kappa_mu_shadowed_HOS}
For the high- and low-SNR regimes, the higher-order capacity statistics of spectrum aggregation systems over correlated $\kappa$-$\mu$ shadowed fading channels can be expressed as
\begin{align}
\Lambda _n^\infty &= \frac{A}{{{\ln }^n}2}{\left( {\frac{\eta }{{\bar \gamma }}} \right)^U}\sum\limits_{k = 0}^\infty  {{D_k}} {\sum\limits_{q = 0}^\infty  {\frac{{{{\left( {mM\!+\!k} \right)}_q}}}{{{{\left( U \right)}_q}q!}}\left( {\frac{\eta }{{\bar \gamma \left( {1 \!+\! {\lambda ^{ - 1}}} \right)}}} \right)} ^q}\notag \\
&\times Q\left( {q + U-1,\frac{\eta }{{\bar \gamma }},n} \right),\label{eq:high_cor_kappa_mu_shadowed_HOS} \\
\Lambda _n^{\gamma  \to 0} &= An!\sum\limits_{k = 0}^\infty  {\sum\limits_{p = 0}^\infty  {\frac{{S_{p + n}^n{D_k}}}{{\left( {p + n} \right)!}}} } \frac{{\Gamma \left( {p + n + \mu M} \right)}}{{{{\left( {\eta /\bar \gamma } \right)}^{p + n}}}}\notag \\
&\times {}_2{F_1}\left( {mM + k,U + p + n;,U;\frac{1}{{1 + {\lambda ^{ - 1}}}}} \right). \label{eq:low_cor_kappa_mu_shadowed_HOS}
\end{align}
\end{lemm}
\begin{IEEEproof}
The proof concludes by following a similar line of reasoning as in Lemma \ref{lemm:high_iid_kappa_mu_shadowed_HOS}.
\end{IEEEproof}
According to Lemma \ref{lemm:high_cor_kappa_mu_shadowed_HOS}, a higher $\lambda$ increases the HOS of the channel capacity. This is anticipated, since larger values of $\lambda$ reduce the correlation between frequency bands, making receiver signal more stronger.

{\subsection{Practical Implementation of HOS}
To evaluate the performance of spectrum aggregation systems, several important measures will in discussed by using the HOS of the channel capacity presented above. These measures can also serve as useful tools for the design of practical dispersed spectrum cognitive radio (DS-CR) systems. First of all, the amount of fading (AoF) of the channel capacity or the so called fading figure is defined as the ratio of variance to the square ergodic capacity as $AoF = \frac{{{\Lambda _2}}}{{\Lambda _1^2}} - 1$ \cite{sagias2011higher}. The variance of the channel capacity is denoted by ${\mathop{\rm var}}  = {\Lambda _2} - \Lambda _1^2$ to describe how far the channel capacity lies from the ergodic capacity. Its normalization with respect to the ergodic capacity, $AoD = \frac{{{\Lambda _2} - \Lambda _1^2}}{{{\Lambda _1}}}$, is called the amount of dispersion (AoD). Furthermore, we can define the reliability percentage of the signal throughput as $R=100(1-AoD)$. For good channel quality, AoD approaches to zero and $R \to 100$. In addition, the skewness is a metric of the degree of asymmetry for the distribution of the channel capacity as $ S = \frac{{{\Lambda _3} - \Lambda _1^3}}{{\sqrt {{{\left( {{\Lambda _2} - \Lambda _1^2} \right)}^3}} }} $. For symmetric distributions, $S = 0$, while $S<0$ denotes the distribution is skewed to the left. In addition, the kurtosis corresponds to the degree of peakedness of the channel capacity around the ergodic capacity as $K = \frac{{{\Lambda _4} - \Lambda _1^4}}{{{{\left( {{\Lambda _2} - \Lambda _1^2} \right)}^2}}}$. Within this context, it is worth mentioning that these important statistical metrics of the channel capacity can be also efficiently and accurately computed by using the HOS expressions.}

\section{Numerical Results}\label{se:num}
In this section, we present various performance evaluation results using the HOS of the channel capacity expressions presented in Sections \ref{se:HOS} for spectrum aggregation systems operating over $\kappa$-$\mu$ and $\kappa$-$\mu$ shadowed fading channels, respectively. To validate the accuracy of the aforementioned expressions, comparisons with complementary Monte-Carlo simulated results with $10^6$ realizations of random variables are also included in these figures. We use the approaches presented in \cite{peppas2012sum} and \cite{moreno2016kappa} to generate random variables from the squared $\kappa$-$\mu$ and $\kappa$-$\mu$ shadowed distributions, respectively. The impact of system and fading parameters on the HOS performance of spectrum aggregation systems are discussed in detailed.

\subsection{Convergence of Derived Results}
Since the derived results are given in sum of infinite series, we prove the convergence of the derived results by truncating the appropriate series expressions to achieve an accuracy up to the fifth-significant digit (e.g., ${P_e} \le {10^{ - 5}}$). Table \ref{table1} investigates the impact of the number of moments and fading parameters $\kappa$ and $\mu$ on the convergence of the HOS of the channel capacity for $M=3$. It can be seen from Table \ref{table1} that all infinite series rapidly converged with the speed of convergence for the scenarios of interest. Moreover, the number of terms increases with increasing $\kappa$, $\mu$ and average SNR $\Omega$, while $\mu$ has a noticeable impact on the convergence. For high-SNR regimes, the required number of terms for ergodic capacity ($n=1$) is less than the case of the 4-th statistical moment. However, only a relatively small number of terms is required for the desired accuracy. For the worst case of $n=4$, $\Omega=10$ dB, $\kappa=1$, and $\mu=2$, the maximum number of terms is 23.

{For correlated $\kappa$-$\mu$ shadowed fading channels, the effect of correlation coefficient $\rho$ on the convergence has been studied in Table \ref{table2}. It is clear that the derived results are rapidly converged with less than 40 terms of infinite series. Note that as $\rho$ reduces, the required number of terms decreases. Furthermore, the computation time of derived analytical results is much less than the one of simulations. For example, we spend less than 20 seconds to calculate \eqref{eq:iid_kappa_mu_HOS} by using MATHEMATICA, while the simulation costs more than 230 seconds to derive the same result. Note that other cases have similar fact of converging steadily and rapidly, and requiring little computational effort, which are validated by our conducted numerical experiments.}

\begin{table}[tb]
\renewcommand{\thetable}{\Roman{table}}
\caption{Number of Required Terms for Convergence of the HOS of The Channel Capacity for Spectrum Aggregation Systems over i.i.d. $\kappa$-$\mu$ Fading Channels with ${P_e} \le {10^{ - 5}}$, and $M=3$}
\label{table1}
\centering
\begin{tabular}{|c|c|c|c|c|c|c|}
\hline
\multirow{3}{*}{SNR } &
\multicolumn{3}{c|}{$n=1$} &
\multicolumn{3}{c|}{$n=4$} \\
\cline{2-7}
  & $\kappa=1$& $\kappa=1$ & $\kappa=2$& $\kappa=1$ & $\kappa=1$& $\kappa=2$\\
  & $\mu=1$ & $\mu=2$ & $\mu=1$ &  $\mu=1$ &  $\mu=2$ &$\mu=1$\\
\hline
-10 & 11 & 18 & 17 & 11 & 20 & 16 \\
\hline
0 & 12 & 19 & 18 & 15 & 21 & 20 \\
\hline
10 & 13 & 19 & 18 & 16 & 23 & 21 \\
\hline
\end{tabular}
\end{table}

\begin{table}[tb]
\renewcommand{\thetable}{\Roman{table}}

{\caption{Number of Required Terms for Convergence of the HOS of The Channel Capacity for Spectrum Aggregation Systems over correlated $\kappa$-$\mu$ Shadowed Fading Channels with ${P_e} \le {10^{ - 3}}$, $M=2$, $\kappa_1=1$, $\kappa_2=5$, $\mu_1=1$, $\mu_2=2$, and $m=1$.}}
\label{table2}
\centering
\begin{tabular}{|c|c|c|c|c|c|c|}
\hline
\multirow{2}{*}{SNR} &
\multicolumn{3}{c|}{$n=1$} &
\multicolumn{3}{c|}{$n=2$} \\
\cline{2-7}
& \footnotesize{$\rho \!=\!0.9$} & \footnotesize{$\rho \!=\!0.5$} & \footnotesize{$\rho \!=\!0.1$} & \footnotesize{$\rho \!=\!0.9$} & \footnotesize{$\rho \!=\!0.5$} & \footnotesize{$\rho \!=\!0.1$} \\
\hline
-10 & 35 & 20 & 18 & 38 & 36 & 30 \\
\hline
0 & 30 & 16 & 15 & 33 & 32 & 24 \\
\hline
10 & 26 & 13 & 12 & 29 & 27 & 20 \\
\hline
\end{tabular}
\end{table}

\subsection{Performance Analysis}
\begin{figure}[t]
\centering
\includegraphics[scale=0.6]{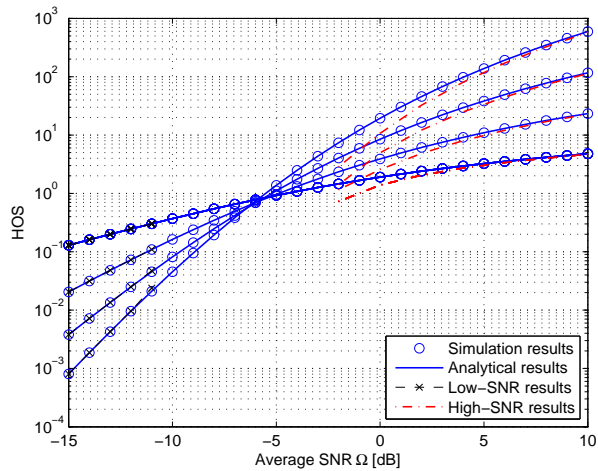}
\caption{Simulated, analytical, and asymptotic HOS of the channel capacity against the average SNR for spectrum aggregation systems over i.i.d. $\kappa$-$\mu$ fading channels ($\kappa=1$, $\mu=1$, and $M=3$).
\label{fig:Theorem1}}
\end{figure}
For spectrum aggregation systems over i.i.d. $\kappa$-$\mu$ fading channels, the simulated, analytical \eqref{eq:iid_kappa_mu_HOS}, and asymptotic \eqref{eq:high_iid_kappa_mu_HOS}, \eqref{eq:low_iid_kappa_mu_HOS} HOS of the channel capacity are plotted against the average SNR $\Omega$ in Fig. \ref{fig:Theorem1}. As seen, the analytical results perfectly match the Monte-Carlo simulations. Clearly, the high-SNR approximations are sufficiently tight and become exact even at moderate SNR values, while a precise agreement between the exact and low-SNR results can be observed. This implies that they can efficiently predict the HOS of the channel capacity over a wide SNR range. Moreover, the gap between the corresponding curves increases as $n$ increases which means the high-SNR approximation is more accurate for low order statistics (e.g., ergodic capacity). It is also clear from Fig. \ref{fig:Theorem1} that the HOS curves get closer to each other almost around -6 dB, which determines the boundary of the high- and low-SNR regimes. Base on the interesting finding of $\Lambda _n=1$ at around -6 dB, we can simply model the HOS of the channel capacity as $\Lambda _n \approx \Lambda _1^n$. Due to the $n$th power of the ergodic capacity, the behavior of the HOS of capacity is different in the high- and low-SNR regimes, respectively. Furthermore, the crossing point will be shifted toward left, and therefore, the ergodic capacity increasing if increasing the values of fading parameters (e.g., $\kappa$ and $\mu$) of each frequency band.

\begin{figure}[t]
\centering
\includegraphics[scale=0.6]{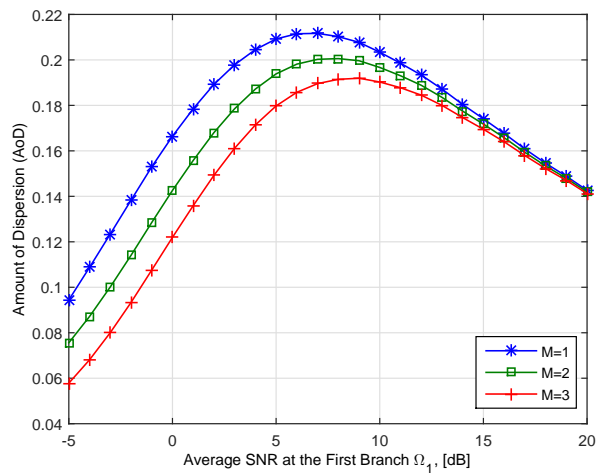}
\caption{Amount of dispersion of the channel capacity against the average SNR for spectrum aggregation systems over i.n.i.d. $\kappa$-$\mu$ fading channels ($\Omega_2=\Omega_3=1$ dB, $\kappa_1=2.5$, $\kappa_2=3.5$, $\kappa_3=4.75$, and $\mu_1=1$, $\mu_2=1$, $\mu_3=2$).
\label{fig:inid_ku_AOD}}
\end{figure}
The effect of the number of frequency bands $M$ on the AoD performance of spectrum aggregation systems over i.n.i.d. $\kappa$-$\mu$ fading channels is shown in Fig. \ref{fig:inid_ku_AOD}.
One can notice that the AoD appears to increase for low and moderate SNRs, while it begins to decrease for the high-SNR regime for all cases. Furthermore, it can be seen from the Fig. \ref{fig:inid_ku_AOD} that the AoD plot becomes peaky at around 9 dB for $M=3$, while the AoD reaches its highest value around 6 dB for $M=1$. With respect to the reliability percentage of the spectrum aggregation system, the transmit SNR should be chosen greater than the SNR for which the AoD peaks. For example, for the case of $M=3$, the maximum AoD is 0.1919 and the reliability percentage is 90.81\%, which means that the average SNR must be chosen equal to or greater than 9 dB in order to reach at least 90.81\% reliable transmission. Moreover, the gap between different number of bands $M$ decreases at high SNRs which implies that the fading effect becomes less pronounced as anticipated.

\begin{figure}[t]
\centering
\includegraphics[scale=0.65]{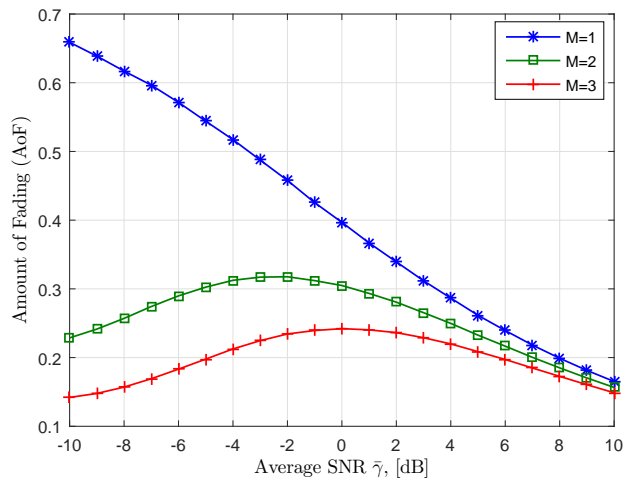}
\caption{Amount of fading of the channel capacity against the average SNR for spectrum aggregation systems over i.i.d. $\kappa$-$\mu$ shadowed fading channels ($\kappa=\mu=2$ and $m=1$).
\label{fig:iid_kum_AOF_M}}
\end{figure}
Figure \ref{fig:iid_kum_AOF_M} depicts the AoF of the channel capacity for spectrum aggregation systems over i.i.d. $\kappa$-$\mu$ shadowed fading channels as a function of average SNR per band $\bar \gamma$ for different sets of frequency bands $M$. It is clear that the AoF decreases drastically as the value of $\bar \gamma$ increases. At low SNRs, the AoF performance of the spectrum aggregation system is significantly improved with increasing the value of frequency bands and $M$. For example, the AoF is 0.665 for the case of $M=1$ at -10 dB, while it reduces to 0.148 for $M=3$. Therefore, we can use more frequency bands to combat the low SNRs of fading channels.

\begin{figure}[t]
\centering
\includegraphics[scale=0.6]{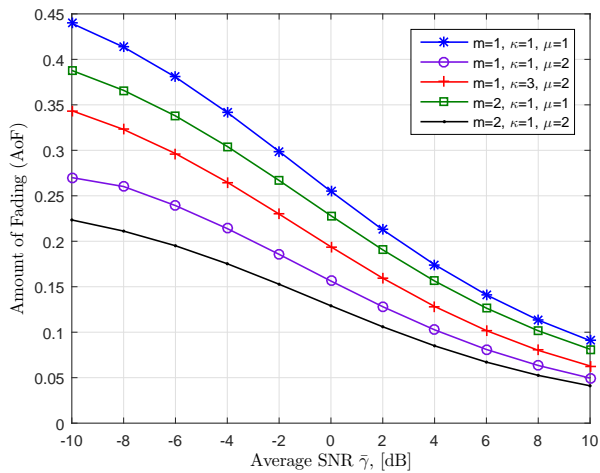}
\caption{Amount of fading of the channel capacity against the average SNR for spectrum aggregation systems over i.i.d. $\kappa$-$\mu$ shadowed fading channels ($M=2$).
\label{iid_kum_AoF_kum}}
\end{figure}
The effect of fading parameters $\kappa$, $\mu$ and $m$ on the AoF of the channel capacity for spectrum aggregation systems over i.i.d. $\kappa$-$\mu$ shadowed fading channels are further investigated in Fig. \ref{iid_kum_AoF_kum}. As indicated by analysis in Section \ref{se:HOS}, the AoF decreases with a smaller value of $\kappa$ (more power of LoS components) and a higher value of $\mu$ (more power of clusters), where the fading channel becomes less deterministic. This finding reveals that more scattered waves are beneficial for improved AoF. One can also notice the increase of the AoF can be obtained for decreasing the shadowing parameter $m$, especially for low SNRs. For example, the value of AoF is 0.275 for the case of $m=1$, $\kappa=1$, $\mu=2$ and $\bar \gamma=-10$ dB, while it reduces to 0.228 for the case of $m=2$, $\kappa=1$, $\mu=2$ and $\bar \gamma=-10$ dB. The impact of fading appears to be particularly critical for low SNRs, while in the high-SNR regime its impact is relatively reduced.

\begin{figure}[t]
\centering
\includegraphics[scale=0.6]{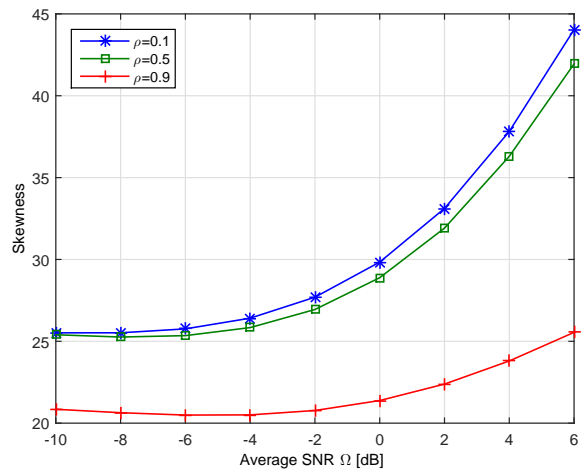}
\caption{Skewness of the channel capacity against the average SNR for spectrum aggregation systems over correlated $\kappa$-$\mu$ shadowed fading channels ($M=3$, $\kappa_i=1$, $\mu_i=2$, and $m=1$).
\label{figcor_kum_Skewness}}
\end{figure}

\begin{figure}[t]
\centering
\includegraphics[scale=0.55]{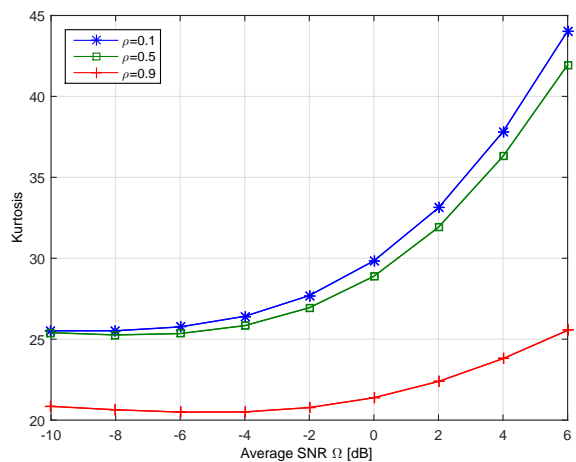}
\caption{Kurtosis of the channel capacity against the average SNR for spectrum aggregation systems over correlated $\kappa$-$\mu$ shadowed fading channels ($M=3$, $\kappa_i=1$, $\mu_i=2$, and $m=1$).
\label{fig:cor_kum_Kurtosis}}
\end{figure}
In Figs. \ref{figcor_kum_Skewness} and \ref{fig:cor_kum_Kurtosis}, Skewness and Kurtosis statistics are plotted against the average SNR $\Omega$, respectively. We consider correlated $\kappa$-$\mu$ shadowed fading channels with three exponential correlation models $\rho_{pq}=\rho^{|p-q|}$, where $\rho=0.1, 0.5, 0.9$. It is clear to see from Figs. \ref{figcor_kum_Skewness} and \ref{fig:cor_kum_Kurtosis} that the skewness and the kurtosis increases as average SNR of each frequency band $\Omega$ increases and/or $\rho$ decreases, showing that the pdf of the channel capacity becomes more spiky with heavy tails and asymmetric. More importantly, the gap between the corresponding curves decreases as $\rho$ decreases which implies that its effect becomes less pronounced.

\section{Conclusion}\label{se:con}
In this paper, we investigate the performance of spectrum aggregation systems over generalized fading channels. In particular, we consider two recently proposed generalized fading models, namely $\kappa$-$\mu$ and $\kappa$-$\mu$ shadowed, which can model propagation phenomena involving LoS and composite fading environments, respectively. Novel and exact expressions for the HOS of the channel capacity of spectrum aggregation systems are derived. Our derived expressions can extend and complement existing results on classical fading models. Furthermore, we deduce simple HOS expressions for the asymptotically low- and high-SNR regimes. Note that all infinite series can be computationally efficient, accurate, and requires only a relative small number of terms for yielding accurate results. Important performance metrics, such as ergodic capacity, variance, AoF, AoD, skewness, and kurtosis, are also derived to show the effects of system and fading parameters on spectrum aggregation systems.
Finally, extensive Monte-Carlo simulations verify the accuracy of the analytical expressions and the tightness of the high- and low-SNR bounds.
The proposed analysis is useful for the spectrum aggregation system design engineer for performance evaluation purposes.

\section*{Appendix}\label{se:app}
\subsection{A Useful Integral Identity}\label{se:integral}
Let us consider an integral in the form of
\begin{align}\label{eq:theta}
{\Theta _\delta }\left( {a,b} \right) = \int_0^\infty  {{{\left( {1 + x} \right)}^\delta }{x^{a - 1}}{e^{ - bx}}dx},
\end{align}
where $a \in \mathbb{N}$, $b>0$, and $\delta \in \mathbb{C}$. With the help of \cite[Eq. (39)]{kang2006capacity}, \eqref{eq:theta} can be also expressed as
\begin{align}
&{\Theta _\delta }\left( {a,b} \right)= \Gamma \left( a \right)U\left( {a,a + \delta  + 1;b} \right)\notag \\
&={e^b}\sum\limits_{k = 0}^{a - 1} {\left( {\begin{array}{*{20}{c}}
{a - 1}\\
k
\end{array}} \right)} \frac{{{{\left( { - 1} \right)}^{a - k - 1}}}}{{{b^{\delta  + k + 1}}}}\Gamma \left( {\delta  + k + 1,b} \right)\label{eq:theta_result}
\end{align}
where $U(\cdot)$ denotes the Tricomi hypergeometric function \cite[Eq. (13.1.3)]{abramowitz1964handbook}, and $\Gamma(\cdot,\cdot)$ is the upper complementary incomplete gamma function \cite[Eq. (8.350.2)]{gradshtein2000table}.
By using Leibniz's rule \cite{gradshtein2000table}, the $n$th order derivative of \eqref{eq:theta_result} can be evaluated as
\begin{align}
&{\left. {\frac{{{d^n}{\Theta _\delta }\left( {a,b} \right)}}{{d{\alpha ^n}}}} \right|_{\alpha  = \delta  + k + 1}}= {e^b}\sum\limits_{k = 0}^{a - 1} {\left( {\begin{array}{*{20}{c}}
{a - 1}\\
k
\end{array}} \right)} \frac{{{{\left( { - 1} \right)}^{a - k - 1}}}}{{{b^{\delta  + k + 1}}}}\notag \\
&\times \sum\limits_{p = 0}^n {\left( {\begin{array}{*{20}{c}}
n\\
p
\end{array}} \right){{\ln }^{n - p}}\left( {\frac{1}{b}} \right)\left\{ {{{\left. {\frac{{{d^n}\Gamma \left( {\alpha ,b} \right)}}{{d{\alpha ^n}}}} \right|}_{\alpha  = \delta  + k + 1}}} \right\}} \label{eq:theta_result_1} \\
&= n!{e^b}\sum\limits_{k = 0}^{a - 1} {\left( {\begin{array}{*{20}{c}}
{a - 1}\\
k
\end{array}} \right)} \frac{{{{\left( { - 1} \right)}^{a - k - 1}}}}{{{b^{\delta  + k + 1}}}}\notag \\
&\times G_{n + 1,n + 2}^{n + 2,0}\left( {b\left| {\begin{array}{*{20}{c}}
{\overbrace {1,1, \cdots ,1}^{n + 1\;1's}}\\
{\underbrace {0,0, \cdots ,0}_{n + 1\;0's},\delta  + k}
\end{array}} \right.} \right),\label{eq:theta_result_2}
\end{align}
From \eqref{eq:theta_result_1} to \eqref{eq:theta_result_2}, we have used the differentiation identity \cite[Eq. (06.06.20.0013.01)]{Wolfram2011function} and performed some algebraic manipulations.

Based on \eqref{eq:theta}, the $n$th order derivative of ${\Theta _\delta }\left( {a,b} \right)$ can alternatively be given by
\begin{align}\label{eq:J}
{J_\delta }\left( {a,b,n} \right) = \int_0^\infty  {{{\left( {1 + x} \right)}^\delta }{{\ln }^n}\left( {1 + x} \right){x^{a - 1}}{e^{ - bx}}dx}.
\end{align}

By substituting $\delta=1$ into \eqref{eq:theta_result_2}, we can derive the auxiliary function $J\left( {a,b,n} \right)$ as
\begin{align}\label{eq:J_result}
J\left( {a,b,n} \right)&= n!{e^b}\sum\limits_{k = 0}^{a - 1} \Bigg[ {{\left( { - 1} \right)}^{a - k - 1}}\left( {\begin{array}{*{20}{c}}
{a - 1}\\
k
\end{array}} \right){b^{ - 1 - k}}\notag \\
&\times G_{n + 1,n + 2}^{n + 2,0}\left( { b \left|\begin{array}{*{20}{c}}
{\overbrace {1,1, \cdots ,1}^{n + 1\;1's}}\\
{\underbrace {0,0, \cdots ,0}_{n + 1\;0's},1 + k}
\end{array}\right.} \right) \Bigg].
\end{align}

\subsection{High-Order Differentiation}\label{se:differ}
With the help of Leibniz's rule \cite{gradshtein2000table}, the $n$th differentiation of the product of the gamma function and the power functions can be expressed as
\begin{align}\label{eq:Q_result_1}
&Q\left( {a,b,n} \right)=\frac{{{d^n}}}{{d{a^n}}}\left( {\frac{{\Gamma \left( {a + 1} \right)}}{{{b^{a + 1}}}}} \right) \notag \\
& =\sum\limits_{k = 0}^n {\left( {\begin{array}{*{20}{c}}
n\\
k
\end{array}} \right)} \frac{{{{\left( { - 1} \right)}^{n - k}}{{\ln }^{n - k}}\left( b \right)}}{{{b^{a + 1}}}}\frac{{{d^k}\Gamma \left( {a + 1} \right)}}{{d{a^k}}}.
\end{align}
Utilizing the high-order differentiation of the gamma function \cite[Eq. (10)]{yilmaz2012novel}, we can derive
\begin{align}\label{eq:Q_result}
&Q\left( {a,b,n} \right)= \sum\limits_{k = 0}^n {\left( {\begin{array}{*{20}{c}}
n\\
k
\end{array}} \right)\frac{{{{\left( { - 1} \right)}^{n - k}}{{\ln }^{n - k}}\left( b \right)k!}}{{{b^{a + 1}}}}} \notag \\
&\times \Bigg( G_{k + 1,k + 2}^{k + 2,0}\left( {1\left| {\begin{array}{*{20}{c}}
{\overbrace {1,1, \cdots ,1}^{k + 1\;1's}}\\
{1 + a,\underbrace {0,0, \cdots ,0}_{k + 1\;0's}}
\end{array}} \right.} \right) \notag \\
&+ {{\left( { - 1} \right)}^k}G_{k + 1,k + 2}^{1,k + 1}\left( {1\left| {\begin{array}{*{20}{c}}
{\overbrace {1,1, \cdots ,1}^{k + 1\;1's}}\\
{1 + a,\underbrace {0,0, \cdots ,0}_{k + 1\;0's}}
\end{array}} \right.} \right) \Bigg).
\end{align}

\bibliographystyle{IEEEtran}
\bibliography{IEEEabrv,Ref}

\end{document}